\documentclass[10pt, twocolumn, floatfix, notitlepage, superscriptaddress]{revtex4-1}

\usepackage[tophrase=\,--\,, range-units=single, multi-part-units=single, product-units=single, separate-uncertainty=true]{siunitx}
\usepackage{graphicx}
\usepackage{miller} 
\usepackage[utf8]{inputenc}

\newcommand{\s}[1]{\mathrm{#1}} 

\newlength{\figwidth}
\newlength{\smallfigwidth}
\newlength{\fullfigwidth}
\newlength{\smalltwocolfigwidth}
\begin{document}

\title{Surface Acoustic wave modulation of a coherently driven quantum dot in a pillar microcavity}

\author{B. Villa}
\email{bruno.villa@crl.toshiba.co.uk}
\affiliation{Toshiba Research Europe Limited, Cambridge Research Laboratory, 208 Cambridge Science Park, Milton Road, Cambridge CB4 0GZ, United Kingdom}%
\affiliation{Cavendish Laboratory, University of Cambridge, J. J. Thomson Avenue, Cambridge CB3 0HE, United Kingdom}
\author{A. J. Bennett}
\author{D. J. P. Ellis}
\affiliation{Toshiba Research Europe Limited, Cambridge Research Laboratory, 208 Cambridge Science Park, Milton Road, Cambridge CB4 0GZ, United Kingdom}%
\author{J. P. Lee}
\affiliation{Toshiba Research Europe Limited, Cambridge Research Laboratory, 208 Cambridge Science Park, Milton Road, Cambridge CB4 0GZ, United Kingdom}%
\affiliation{Engineering Department, University of Cambridge, 9 J. J. Thomson Avenue, Cambridge, CB3 0FA, United Kingdom}
\author{J. Skiba-Szymanska}
\affiliation{Toshiba Research Europe Limited, Cambridge Research Laboratory, 208 Cambridge Science Park, Milton Road, Cambridge CB4 0GZ, United Kingdom}%
\author{T. A. Mitchell}
\author{J. P. Griffiths}
\affiliation{Cavendish Laboratory, University of Cambridge, J. J. Thomson Avenue, Cambridge CB3 0HE, United Kingdom}
\author{I. Farrer}
\altaffiliation{Present address: Department of Electronic \& Electrical Engineering, University of Sheffield, Mappin Street, Sheffield S1 3JD, United Kingdom}
\affiliation{Cavendish Laboratory, University of Cambridge, J. J. Thomson Avenue, Cambridge CB3 0HE, United Kingdom}
\author{D. A. Ritchie}
\author{C. J. B. Ford}
\affiliation{Cavendish Laboratory, University of Cambridge, J. J. Thomson Avenue, Cambridge CB3 0HE, United Kingdom}
\author{A. J. Shields}
\affiliation{Toshiba Research Europe Limited, Cambridge Research Laboratory, 208 Cambridge Science Park, Milton Road, Cambridge CB4 0GZ, United Kingdom}%

\date{\today}%

\begin{abstract}
	We report the efficient coherent photon scattering from a semiconductor quantum dot embedded in a pillar microcavity. We show that a surface acoustic wave can periodically modulate the energy levels of the quantum dot, but has a negligible effect on the cavity mode. The scattered narrow-band laser is converted to a pulsed single-photon stream, displaying an anti-bunching dip characteristic of single-photon emission. Multiple phonon sidebands are resolved in the emission spectrum, due to the absorption and emission of vibrational quanta in each scattering event.
\end{abstract}


\maketitle


\section{Introduction}

	Single-photon sources (SPSs) are a key component in emerging quantum technologies \cite{gisin_quantum_2002,waks_secure_2002,shields_semiconductor_2007,aharonovich_solid-state_2016}. Semiconductor quantum dots (QDs) are one of the most technologically advanced candidates for a practical SPS due to their high internal quantum efficiency \cite{aharonovich_solid-state_2016}, narrow linewidths \cite{kuhlmann_transform-limited_2015} and tunability. A major area of study has been the engineering of microfabricated photonic structures around the dots to increase the efficiency with which photons are collected \cite{lodahl_interfacing_2015}. Most promising are those cavities which both enhance the emission rate and funnel photons into a Gaussian mode that can be efficiently collected. Nanowire antennae can achieve the highest collection efficiencies \cite{claudon_highly_2010} but display no Purcell enhancement. Conversely, photonic crystals can obtain the highest Purcell enhancements but only by suppressing the easy-to-collect vertical emission \cite{englund_controlling_2005}. In contrast pillar microcavities can achieve state-of-the-art performance in both metrics. Recently, there has been renewed interest in these cavities as it has been shown that, in conjunction with resonant excitation, they can act as photon sources with unprecedented efficiency, purity and indistinguishability \cite{he_-demand_2013,somaschi_near-optimal_2016,bennett_cavity-enhanced_2016,wang_near-transform-limited_2016}. 

	\begin{figure}[b]%
		\centering
		\includegraphics[width=\smallfigwidth]{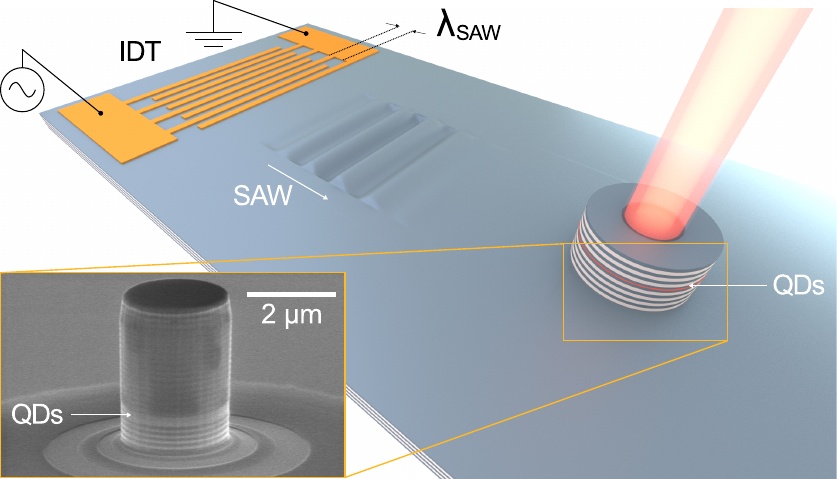}
		\caption{Schematic sample design. The device consists of an interdigitated transducer (IDT) and an optical cavity in the path of the SAW. QDs in the cavity are excited optically. Inset: scanning electron micrograph of a pillar microcavity partially etched through the bottom mirror.}
		\label{fig1}
	\end{figure}

	In addition to engineering the photonic density of states it is also possible to modify the phonon environment of a dot through additional phononic engineering \cite{balram_coherent_2016} or by driving the sample with a sound wave. A sound wave excited on the surface of a piezoelectric semiconductor, and known as a surface acoustic wave (SAW), can change the separation of the energy levels in the dot as the lattice is stretched and compressed \cite{white_surface_1970,lima_modulation_2005}. SAWs can be used for high frequency modulation at frequencies of MHz to tens of GHz \cite{tadesse_acousto-optic_2015,kapfinger_dynamic_2015,fuhrmann_dynamic_2011,li_nanophotonic_2015}. Most reports on SAW-modulated quantum structures to date are limited to samples without Purcell effect \cite{gell_modulation_2008,couto_photon_2009,metcalfe_resolved_2010}. Although there is a recent report of SAW-modulated emission from a photonic crystal cavity \cite{weis_surface_2016}, a hybrid platform combining SAWs and pillar microcavities has clear advantages and is yet to be investigated. Here, we show that it is possible to modulate the energy levels of a QD inside a pillar microcavity with a SAW. The cavity mode retains the ability to enhance the emission of the dot and is not modulated by the SAW. We perform resonant excitation of this system, to create a stream of single-photons at a rate defined by the SAW.

	\begin{figure}[!h]%
		\centering%
		\includegraphics[width=0.935\smallfigwidth]{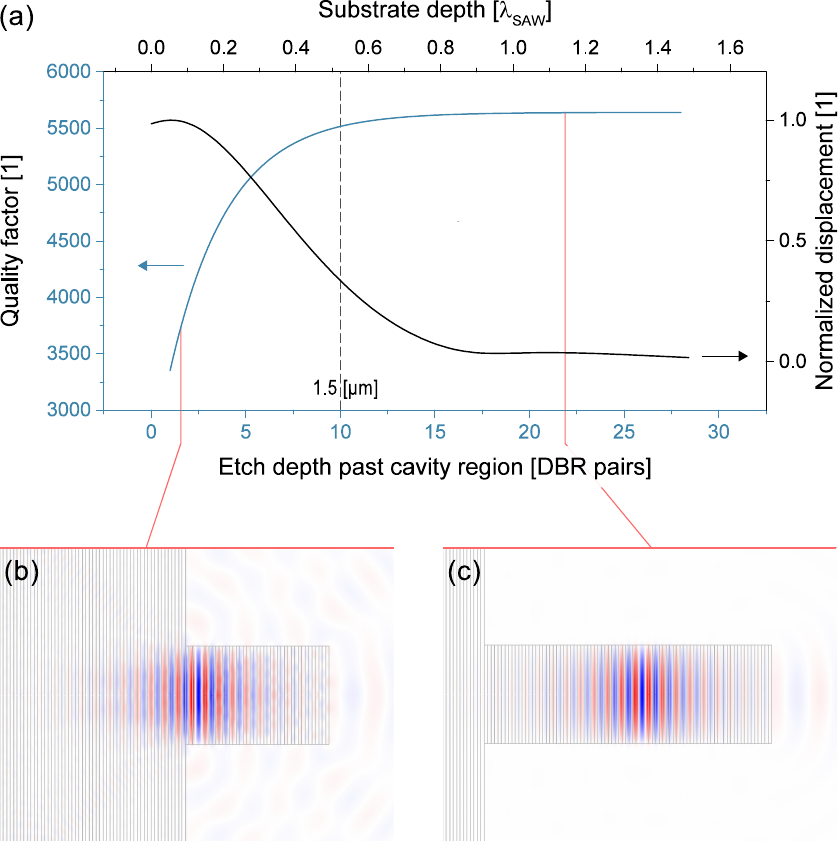}%
		\caption{FEM simulations. (a) Extracted Q factors of pillar microcavities etched a fixed number of DBR pairs past the cavity region (bottom left) and substrate displacement for a \SI{1}{GHz} SAW (top right). Ten DBR periods corespond to \SI{1.5}{\micro m} in this device. (b, c) Out of plane electric field component of the cavity mode at selected etch depths (1, 22 DBR pairs).}%
		\label{fig2}%
	\vspace{2em}%
		\centering%
		\includegraphics[width=\figwidth]{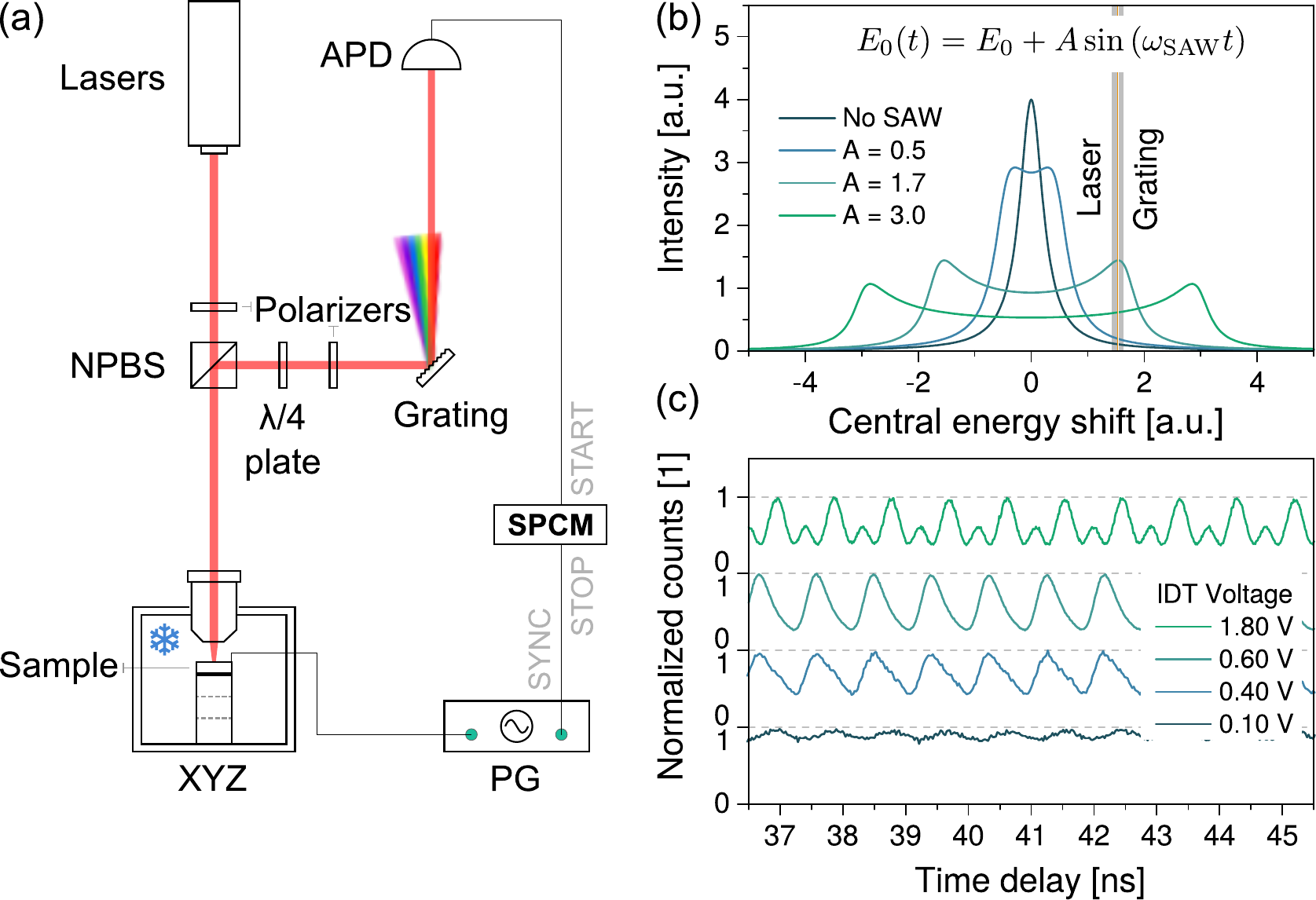}%
		\caption{Time-resolved measurement of the SAW tuning. (a) Experimental set-up around a closed-cycle cryostat (PG: pulse generator, SPCM: single-photon counting module, APD: avalanche photodiode, NPBS: non-polarizing beam splitter). (b) Illustration of the time-averaged spectra for a sinusoidally tuned Lorentzian line as a function of the tuning parameter $A$, as defined in the inset. The position of the detuned excitation laser and the grating window is marked. (c) Time-resolved measurement of the modulation for increasing applied voltage on the input transducer and thus increasing SAW modulation. The curves are offset for clarity. For low modulation amplitudes the transition is not shifted enough to compensate for the initial detuning. As the modulation increases the periodic driving with $f_\s{SAW}$ becomes evident as photons start to be detected. With even stronger tuning the transition overlaps with the resonant laser and grating twice per period and a second peak becomes visible.}%
		\label{fig3}%
	\end{figure}

	The sample used in these experiments is schematically depicted in Fig. \ref{fig1}. The dots are embedded in a distributed Bragg reflector (DBR) optical cavity grown by molecular beam epitaxy. The top and bottom DBRs consist of 18 and 28 repeats of $\lambda/4$ thick alternating GaAs/AlGaAs layers, respectively. The cavity is etched into pillars of \SI{2.1}{\micro\meter} diameter with Q factors around 3000 by conventional dry etching. A single-finger Ti/Au interdigitated transducer is deposited on the etched surface to generate a SAW through the piezoelectric effect. In this simple design, each finger alternates in voltage and is $\lambda_\s{SAW}/4$ wide. The transducer is designed to excite a SAW in GaAs in the \hkl[0 1 -1] direction at $f_\s{SAW}\sim\SI{1}{GHz}$. 

	An important consideration is that the optimal etch depths for the phononic and photonic aspects of the device do not coincide. Because the SAW is confined to within one acoustic wavelength normal to the surface  \cite{morgan2010surface}, the planar cavity should ideally only be etched to the QD growth plane to achieve maximal strain tuning. However, this would compromise the cavity mode confinement and lower the Q factor compared to a pillar that is etched completely through the bottom mirror. Finite element method (FEM) simulations corroborate this intuitive picture, as illustrated in Fig. \ref{fig2}. Both the Q factor and the mode confinement converge quickly towards their achievable limits with increasing etch depth past the cavity region. Thus, it is only necessary to etch the first few periods into the bottom DBR to get a cavity enhancement, while keeping the QDs in range of the SAW. All data shown in the following sections is the result of studies on a charged transition from a single QD in a microcavity etched 7 periods into the DBR. It shows a Purcell enhancement factor $F_\s{P}\sim2$ and a radiative lifetime of $\SI{530}{ps}$.

\section{SAW tuning}

	The charged transition under study is brought close to resonance with the cavity by means of temperature tuning. At \SI{26}{K} the two are close to degeneracy, with the cavity fixed at \SI{891.26}{nm} (\SI{1.391}{eV}). The addition of a SAW to the system completes the resonance tuning through the strain induced modulation. As the SAW passes through the pillar, the central wavelength of the emission follows the sinusoidal mechanical wave. Averaged over time, this process gives rise to a two lobed intensity profile, with the maxima at the modulation edges and a net broadening of the line shape. When sending the collected emission through a monochromating grating, a small SAW-induced tuning suffices to periodically move the central wavelength in and out of the corresponding detection window. It is worth noting that we do not observe signs of simultaneous tuning of the cavity mode by the SAW.

	Since the transition energy is constantly changing due to the SAW, the resonance condition for optical excitation is only met at certain points during the SAW cycle. This phase is controlled by the detuning between the unperturbed transition and the laser as well as the SAW amplitude. For our experiments the resonant excitation is set to happen at one of the modulation extrema, coinciding with the spectral region filtered by the grating. This is achieved using the set-up outlined in Fig. \ref{fig3}(a), where an APD and a single-photon counting card are used to record arrival times of photons relative to the SAW. The light is sent through a grating to remove any non-resonant light, while the resonant laser contribution is suppressed by cross-polarization filtering. Figure \ref{fig3}(b) schematically shows the principle and a series of recorded histograms for increasing SAW amplitude is presented in Fig. \ref{fig3}(c). When the tuning by the SAW is lower than the deliberate detuning of the transition no light reaches the detector, resulting in a flat curve. As the SAW amplitude increases, so does the overlap of the transition with the laser and grating window and a periodic signal starts to appear. Further increment of the tuning range results in the transition crossing this resonance point twice per period and a second peak becomes visible in the trace. We thus set the amplitude such that only one peak is clearly visible in the time-resolved measurement.

\section{Single-photon emission}

\begin{figure}%
		\centering
		\includegraphics[width=\smallfigwidth]{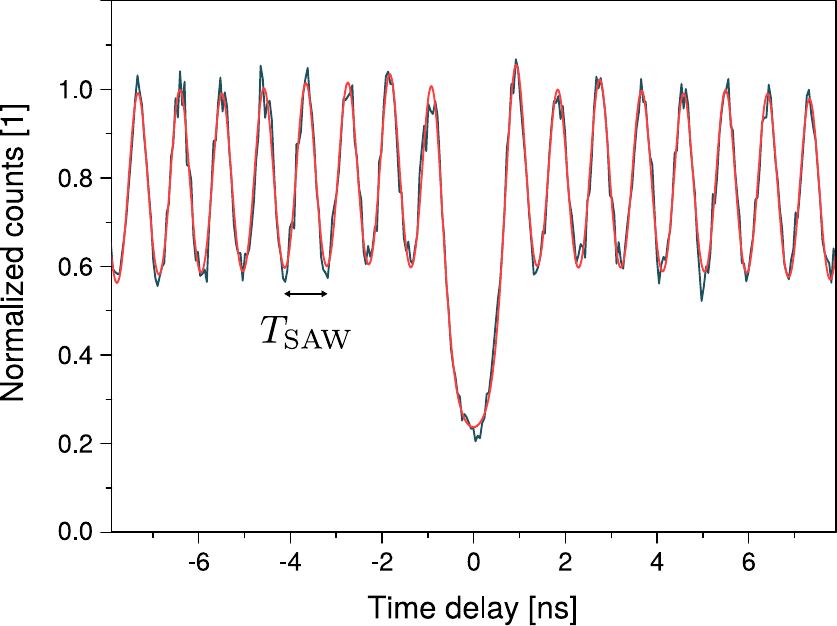}
		\caption{Second order autocorrelation measurement under cw resonant excitation. A continuous SAW is applied such that the blue extreme of the modulation overlaps with the laser and cavity. The pulsed curve reflects the SAW modulation period and shows anti-bunching at zero delay. The red line is a cumulative Voigt peak fit with fixed width.}
		\label{fig4}
	\end{figure}

	In order to prove single-photon emission, the setup from Fig. \ref{fig3}(a) was modified to include a Hanbury-Brown and Twiss interferometer after the grating. Measurements of the second order autocorrelation function $g^{(2)}(\tau)$ were carried out under resonant, continuous wave (cw) optical excitation (\SI{16}{nW} laser power). A small amount (\SI{4}{nW}) of \SI{660}{nm} laser was added in order to generate charge carriers and brighten the trion transition  \cite{bennett_cavity-enhanced_2016}. The SAW was also excited continuously as heating was not an issue. The periodic filtering resulting from the effect of the SAW and the experimental setup gives rise to a pulsed signature in $g^{(2)}(\tau)$, as presented in Fig. \ref{fig4}, where the spacing between pulses corresponds to the SAW period $T_\s{SAW} = \SI{0.92}{ns}$. The central peak is clearly suppressed, with $g^{(2)}(0) =	0.21$ measured in the raw data. It is evident that there is some overlap with the neighbouring peaks, so we can be confident that the actual multi-photon emission in a given pulse is lower. Indeed, a simple fit to the data as a series of overlapping Voigt functions with fixed width yields a curve in reasonable agreement with the measurement (red curve in Fig. \ref{fig4}). An upper bound of $g^{(2)}(0)<\SI{6}{\%}$ can be extracted from the ratio of zero-delay peak area to average peak area at long time delays.

\section{SAW-induced sidebands}
		
\begin{figure}%
		\centering
		\includegraphics[width=\figwidth]{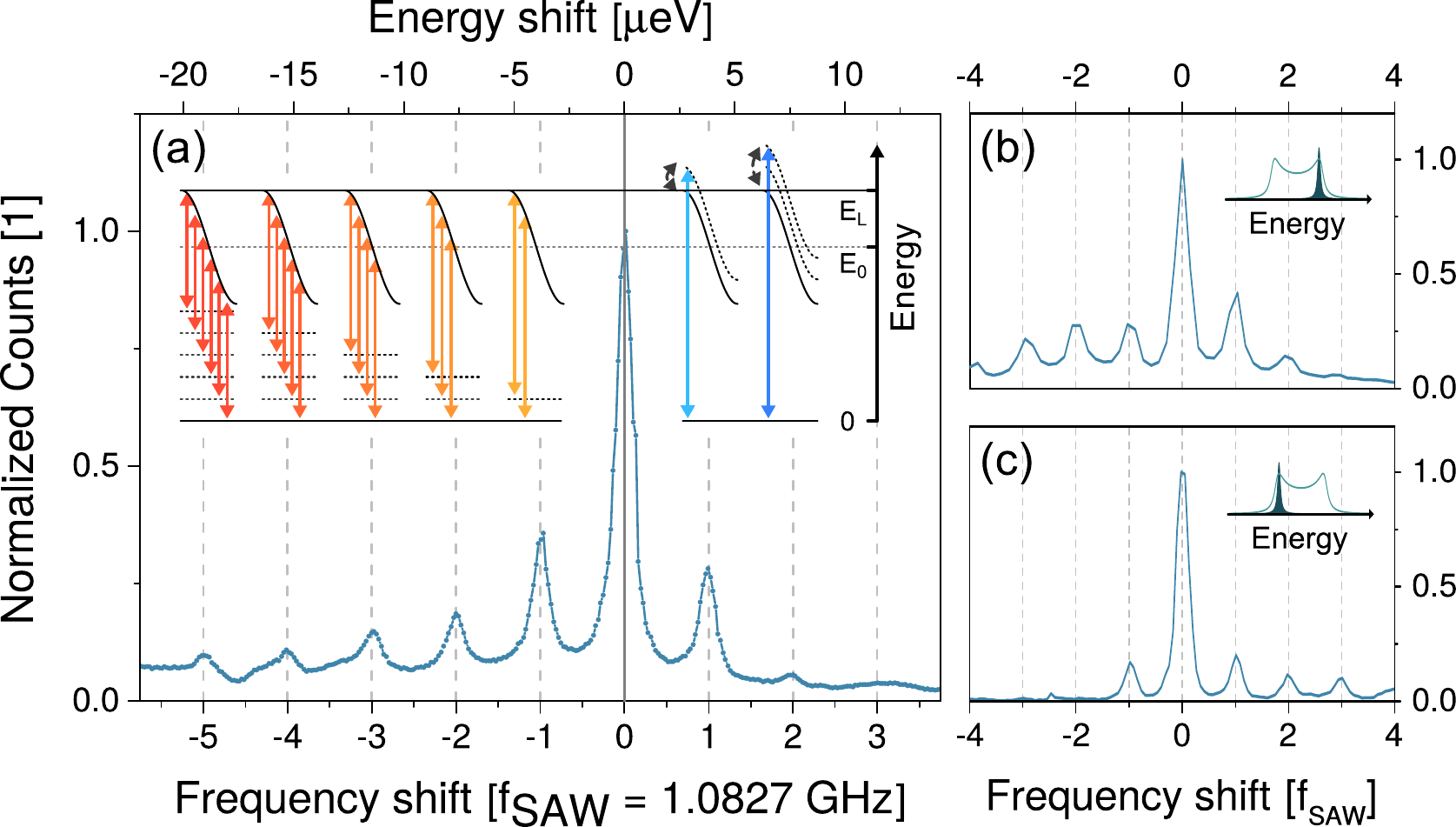}
		\caption{Appearance of sidebands in the presence of a SAW. (a) High resolution spectrum of the resonantly driven transition, showing clear sidebands due to multiple phonon-assisted transitions. The inset schematically shows the energy level diagram and some of the contributing transitions (phonon arrows on the red side are omitted for clarity). The modulation of the transition energy during half a SAW period is shown for each sideband. The energies $E_0$ and $E_\s{L}$ mark the unperturbed transition and laser energy, respectively and the distance between dashed lines represents the phonon energy $\hbar\omega_\s{SAW}$. (b, c) Spectra as in (a) for a the same (b), and opposite (c) spectral alignment of the laser relative to the SAW induced tuning range, as illustrated in the respective insets.}
		\label{fig5}
	\end{figure}

	The simultaneous driving of the transition with a resonant laser and an acoustic field also enables strong phonon-assisted transitions  \cite{metcalfe_resolved_2010}. By introducing a Fabry-Pérot etalon ($\text{free spectral range} = \SI{30}{GHz}$, finesse $\mathcal{F}=156$) between the monochromating grating and the APD the corresponding sidebands could be resolved in our experiments. Fig. \ref{fig5}(a) shows the spectrum resulting from scanning the etalon, while exciting the dot with the resonant (\SI{33.75}{nW}) and red (\SI{4}{nW}) lasers as well as the SAW. In addition to the central carrier line, sidebands at integer multiples of the SAW frequency $f_\s{SAW}=\SI{1.08}{GHz}$ are clearly observed. The asymmetry in the number of sidebands ($>5$ red and 2 blue) is attributed to the position of the laser relative to the transition energy, the reason being that as the central wavelength moves away from the laser, processes involving different numbers of phonons can contribute to the same sideband, as illustrated in the inset. When the laser energy is red-detuned from the unperturbed transition the situation is reversed, showing more blue sidebands, as presented in figures \ref{fig5}(b) and (c).

	Neither the sideband positions nor their relative strength are considerably affected by changes in the resonant laser power over an order of magnitude. The SAW power on the other hand determines the intensity ratio of carrier frequency to sidebands, with the higher SAW amplitude strengthening the sidebands. In all cases, the linewidths of the central peak and sidebands are very close to the instrumental resolution of \SI{192.3}{MHz} (\SI{0.80}{\micro eV}), indicating that at these powers the device operates in the resonant Rayleigh scattering regime  \cite{bennett_cavity-enhanced_2016,loudon2000quantum}. For the measurement of Fig. \ref{fig5} the extracted linewidth of the central peak is \SI{245.3\pm3.7}{MHz} (\SI{1.01\pm0.02}{\micro eV}) and coincides with that of the sidebands within the error margin.

\section{Conclusion}

	We have used an integrated device with surface acoustic waves and cavity-enhanced quantum dots to coherently scatter cw laser light, creating a GHz repetition rate single-photon source. We have shown the scattered light consists of a “comb" of sub-natural linewidth emission peaks, spaced at the SAW frequency. Beyond single-photon generation, the studied surface acoustic wave based platform opens up possibilities for hybrid systems with two coherent manipulation pathways  \cite{balram_coherent_2016,golter_optomechanical_2016} and has the potential to serve as an on-chip quantum interface  \cite{schuetz_universal_2015}.

\begin{acknowledgments}
	This project has received funding from the European Union's Horizon 2020 research and innovation programme under the Marie Sk\l{}odowska-Curie grant agreement No. 642688 (SAWtrain). We also gratefully acknowledge financial support from the EPSRC CDT in Photonic Systems Development and Toshiba Research Europe Ltd.
\end{acknowledgments}


%

\end{document}